# Coherent control in a semiconductor optical amplifier operating at room temperature

A. Capua[1,3], O. Karni[1], G. Eisenstein[1], V. Sichkovskyi[2], V. Ivanov[2], J. P. Reithmaier[2]

[1]*Electrical Engineering Department, Technion, Haifa 32000, Israel*
[2]*Technische Physik, Institute of Nanostructure Technologies and Analytics,CINSaT, University Kassel, Germany*
[3]*Present address: IBM Almaden Research Labs, IBM, San Jose 95120, USA 2010*

*[acapua@us.ibm.com](mailto:acapua@us.ibm.com)

**Abstract:** We demonstrate the Ramsey analogous experiment known as coherent control, taking place along an electrically-driven semiconductor optical amplifier operating at room temperature.

## 1. Introduction

Coherent control has been demonstrated in a variety of quantum systems. In semiconductors however, tailoring the quantum interaction of the resonant interband transition is considered more challenging. Being a densely packed interacting atomic system, semiconductors exhibit extremely short coherence times of no more than a few hundreds of femtoseconds at room temperature.

Recently, we demonstrated an experiment that enables to readily identify quantum coherent interactions taking place during the propagation of a short pulse in a semiconductor optical amplifier (SOA) even in the presence of significant room-temperature dephasing. The principle of the experiment was that the complex electromagnetic field (phase and amplitude) of the pulse was characterized at the SOA output with a temporal resolution of nearly a single femtosecond. This was demonstrated for the first time in a quantum wire-like (commonly referred as quantum dash) SOA [1] where systematic, coherent Rabi oscillations and self-induced transparency were clearly observed. Furthermore, by a comparison to a simulation of the Maxwell and Schrödinger equations [2], we were able to decipher the time evolution of effective ensemble states along the propagation axis. Immediately after the observation of ref. [1], the results were reported in an inhomogeneously broadened quantum dot gain medium [3].

Using the high resolution technique demonstrated in ref. [1] & [3], we turn in the present work to show coherent control of ensemble states taking place along a semiconductor optical amplifier for the first time. This is shown by sending pairs of optical pulses separated by a known time interval and measuring the coherent signatures imprinted on the second (trailing) pulse. Surprisingly, these are observed even under room-temperature conditions and in the presence of an inhomogeneously broadened gain medium.

## 2. Experimental results

The origin of the high temporal resolution of the pump-probe experiment reported in this paper stems from the fact that the actual field is measured rather than the transmission of the probe pulse. This is achieved by utilizing the cross-correlation frequency-resolved optical gating (X-FROG) technique [4]. A schematic of the experimental arrangement is depicted in Fig. 1. 150 fs pulses at a wavelength of 1530 nm, emitted by an optical parametric oscillator (OPO) were split to form two identical pump and probe pulses separated by a predetermined delay, $\Delta\tau$. The two pulses are coupled to a 1.5 mm-long quantum dot optical amplifier, at the output of which the complex electromagnetic field of both pulses was retrieved from the X-FROG measurement having a temporal resolution reaching nearly 1 fs. The active gain region of the device under test comprised four InAs quantum dot layers and was biased to the absorption regime.

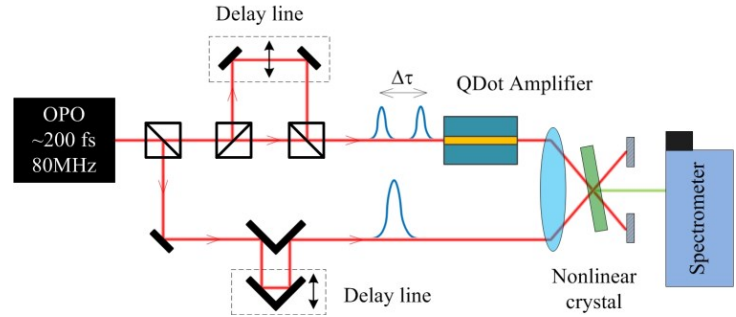

Fig. 1. The pump-probe-FROG characterization setup. Short pulses emitted from an OPO were split into phase-locked pump and probe pulses. After propagating through the InAs QD semiconductor medium, the complete optical field was measured using the X-FROG scheme. The experiments were carried out at room temperature.

The measured optical fields of the pump and probe pulses are presented in Fig 2. The time dependent electrical field, E(t), is represented by the intensity envelope, I(t) : $E(t)=\sqrt{I(t)} \cdot e^{i\varphi(t)}$ with the instantaneous frequency (chirp), ν(t), being the time derivative of the phase, φ(t) : $\nu(t)=-(1/2\pi)(d\varphi/dt)$. We examine traces having a nominal pump-probe delay of 500 fs and longer.

The experiments are performed so that we first select a nominal delay, Δτ, and then scan it in 1 fs steps (which is about one fifth of an optical cycle). For each step, the amplitude and chirp profiles for both pulses are acquired. Exemplary results are shown in Fig. 2a, 2b and 2c for nominal Δτ values of 515, 640 and 900 fs, respectively.

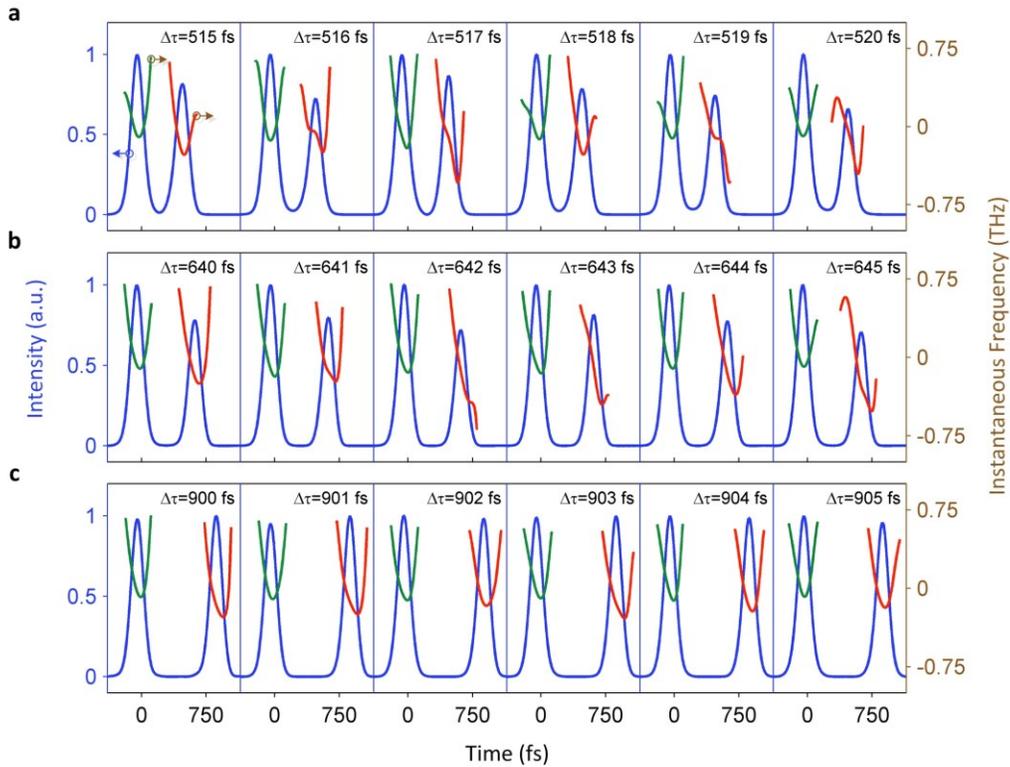

Fig. 2. Measured electromagnetic field. The amplitude (blue traces) and instantaneous frequency (chirp) profiles (green- pump, red- probe) for nominal delays of (a) 515 fs (b) 640 fs and (c) 900 fs as measured by the X-FROG system. Instantaneous frequency traces are measured relative to the carrier frequency. Intensity profiles are normalized to reach unity at their peak.

The amplitude and instantaneous frequency profiles of all leading pump pulses arriving at t=0 are identical since the semiconductor fully decoheres during the time between subsequent pulse pairs. The parabolic chirp profile of the leading pump pulses is a well-known signature of the interaction stemming from a modulation of the refractive index by the excitation of carriers under absorbing conditions. The modulation of the refractive index by the excited population plays a dominant role in understanding the traces obtained by the X-FROG system and is the source of the high sensitivity to the state of the quantum ensemble [1].

When the pump pulse arrives, all quantum states comprising the ensembles are well averaged over the Bloch state-space and hence all pump pulses interact in an identical manner. In contrast, the probe pulses exhibit a fundamentally different behavior. While their amplitude shows some small variations as the delay is scanned, the instantaneous frequency profiles change considerably for every delay step as seen in Fig. 2a for the nominal delay of 515 fs. The figure shows for example that a probe pulse arriving at a delay of 515 fs experiences a valley-like chirp profile similar to that of the pump. However, the phase change associated with a delay that is merely one femtosecond longer, yields a completely different chirp profile. Further tuning of the delay results in a continuous variation of the probe chirp profile. Recalling that the evolution of the quantum states is imprinted on the phase of the electromagnetic field and hence on the chirp profile [1], means that for every delay the probe pulse interacts differently with the ensemble of quantum systems comprising the medium. As the probe pulse is further delayed and the chirp profile continues to vary, a point may be reached where the probe chirp profile can actually change its nature. This is observed in Fig. 2a for the delay of 520 fs where the instantaneous frequency initially increases instead of decreasing suggesting that the leading edge of the probe pulse actually undergoes an amplification process rather than absorption.

The fact that the pump pulse always experiences the same parabolic chirp behavior while the chirp profile of the probe depends on its arrival time implies that the pump pulse locks the phases of the quantum states comprising the ensemble (as predicted in ref. [2]), which otherwise would have been spread randomly over a $2\pi$ range. Once the quantum phases of the ensembles are aligned and well defined, time-keeping becomes possible so that the arrival time of the probe pulse plays a role in determining the outcome of the interaction. Fig. 2a presents only a subset of six traces around the delay of 515 fs but the optical fields behave in the same manner at longer delays. This behavior persists of course as long as the quantum states along the amplifier have not decohered.

For longer nominal delays, the quantum states comprising the ensemble start to dephase and gradually the coherent imprints diminish. The experimental system is capable therefore of measuring directly the evolution of the dephasing process and to determine the time, following the pump perturbation, that quantum coherence ceases to exist.

Fig. 2b illustrates the results measured when the probe is nominally delayed by 640 fs. This subset of measurements shows that clear coherent signatures are still observable however they are less pronounced compared to the traces in Fig. 2a. As the probe pulse is further delayed to a nominal delay of 900 fs, the coherent signatures disappear completely and the profiles of both pump and probe pulses are similar in nature exhibiting a parabolic form independent of the delay. This means that the quantum states have fully dephased and that the electromagnetic field of the second probe pulse interacts with the semiconductor in a similar manner to the first pulse. These experiments constitute therefore a direct measurement of the quantum decoherence.

In order to obtain a physically meaningful figure of merit for the decoherence time, we estimate the time that the quantum states have been given to dephase without the presence of the electromagnetic field which serves as an aligning potential. Limiting the field to a level of 1% of the pulse peak intensity, results in a dephasing time ranging from 360 to 390 fs for the room-temperature semiconductor under the specific bias conditions chosen here. The figure of merit, extracted from the coherent control experiment is similar to coherence times obtained from temperature-dependent photoluminescence measurements of a single quantum dot and from transient four wave mixing measurements in a room-temperature GaAs SOA.

## 3. Numerical calculation

The experimental results can be modeled qualitatively considering that the SOA behaves as an effective ensemble of two-level systems. Describing the semiconductor medium as sequentially placed two-level systems which are fed incoherently from a carrier reservoir has been used to describe propagation of short pulses along a quantum cascade laser26 and in inhomogeneously broadened self-assembled quantum dashes[1] and quantum dots [3]. The model calculates the coevolution of the electromagnetic field and the electronic wavefunctions along the propagation axis for every time iteration according to the coupled Maxwell and Schrödinger equations using the density matrix formalism (see ref. [2]).

A calculated example describing the interaction along the propagation axis at an arbitrary iteration step of the simulation is shown in Fig. 3a. The figure shows the intensity of the pulses together with the spatially dependent population inversion of the effective two-level system, $\rho_{11}$-$\rho_{22}$, where $\rho_{11}$ and $\rho_{22}$ denote the probability amplitudes of the excited and ground states, respectively, and are in fact the diagonal elements of the density matrix. In Fig. 3a, we plot the calculated interaction for two different pump-probe delay intervals, 500 fs and 502 fs, which correspond to a phase difference of almost $\pi$ in the optical field. As in the experiment, the simulation was prepared so that the gain medium is initially in the absorbing regime where the population inversion is negative. For both delay values, the pump pulse undergoes an absorption process which increases the population inversion. The probe arriving at a delay of 500 fs experiences a fully absorptive process (with the population inversion continuously increasing). However, the probe pulse arriving 2 fs later, at 502 fs, experiences an amplification process on its leading edge (with the population inversion decreasing) which turns into an absorption process for the rest of the pulse duration. Since the index of refraction is related to the excited carrier populations, the two cases of absorption and gain mark complimentary signatures in the resultant chirp of the optical field.

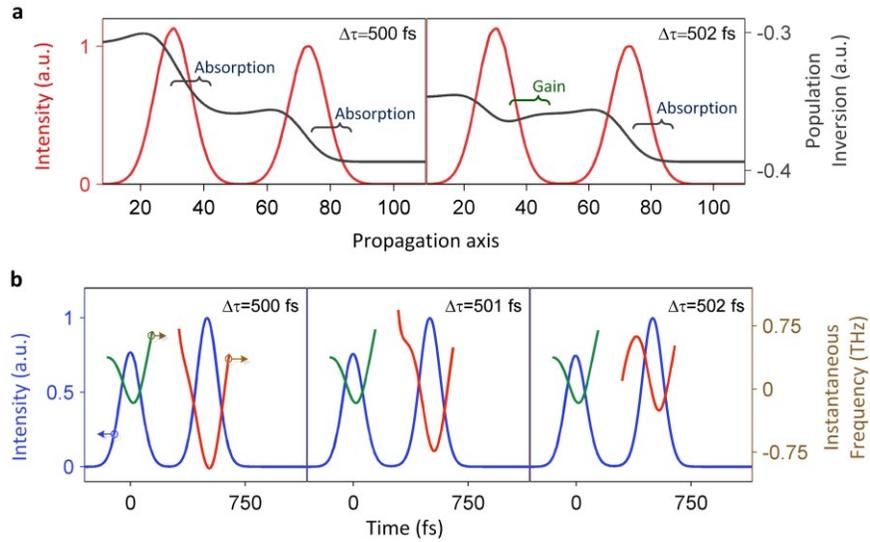

Fig. 3. (a) Calculated example of the intensity (red) and population inversion of the effective two-level system (black) along the propagation axis for two pump-probe delays around a nominal delay of 500 fs. (b) Calculated normalized intensity (blue) and instantaneous frequency (green- pump, red- probe).

Calculated amplitude and chirp temporal profiles of the output pulses are shown in Fig. 3b for three consecutive probe arrival times. All pump pulses are identical and exhibit the parabolic profile, similar to the ones measured experimentally. The two delay intervals, Δτ=500 fs and 502 fs yield a complimentary chirp profile on the leading edge of the probe pulse corresponding to either gain (without inversion) or absorption, consistent with the experimental results of Fig 2a. The calculated response which shows, the initial process of amplification (at Δτ=502 fs), resembles the measured chirp profile for a delay of Δτ=520 fs. As in the experimental results, the simulations reveal that when the probe pulse is further delayed, the coherent features decay and eventually vanish. However, a more quantitative description which fit the experiments requires invoking the inhomogeneous broadened version of the Maxwell-Schrödinger model [3].

**References**

1. A. Capua et al. arXiv:1210.6803 (2012).
2. A. Capua et al. *"A finite-difference time-domain model for quantum-dot lasers and amplifiers in the Maxwell–Schrödinger framework."* IEEE J. Select. Top. Quantum. Electron. **19,** 1900410 (2013).
3. O. Karni et al. *"Rabi oscillations and self-induced transparency in InAs/InP quantum dot semiconductor optical amplifier operating at room temperature."* Opt. Expr. 21, 26786 (2013).
4. R. Trebino, *"Frequency-Resolved Optical Gating: The Measurement Of Ultrashort Laser Pulses"*. (Kluwer Academic Publishers, 2002).